\documentclass[10pt,preprint]{aastex}

\def\ni{\noindent}

\shorttitle{Nonlinear PL relation}
\shortauthors{Ngeow \& Kanbur}

\begin{document}

\title{Nonlinear Period-Luminosity Relation for the Large Magellanic Cloud Cepheids: Myths and Truths}

\author{C. Ngeow}
\affil{Department of Astronomy, University of Illinois, Urbana, IL 61801}

\and

\author{S. M. Kanbur}
\affil{State University of New York at Oswego, Oswego, NY 13126}

\begin{abstract}
In this paper, we discuss and examine various issues concerning the recent findings that suggested the observed period-luminosity (P-L) relation for the Large Magellanic Cloud (LMC) Cepheids is nonlinear. These include (1) visualizing the nonlinear P-L relation; (2) long period Cepheids and sample selection; (3) outlier removal; (4) issues of extinction; (5) nonlinearity of the period-color (P-C) relation; (6) nonlinear P-L relations in different pass-bands; and (7) universality of the P-L relation. Our results imply that a statistical test is needed to detect the nonlinear PL relation. We then show that sample selection, number of long period Cepheids in the sample, outlier removal and extinction errors are unlikely to be responsible for the detection of the nonlinear P-L relation. We also argue for the existence of a nonlinear P-L relation from the perspective of the nonlinear P-C relation and the non-universality of the P-L relation. Combining the evidence and discussion from these aspects, we find that there is a strong indication that the observed LMC P-L relation is indeed nonlinear in the optical bands (however the $K$-band LMC P-L relation is apparently linear). This could be due to the internal physical reasons or the external hidden/additional factors. Compared to the non-linear P-L relation, the systematic error in distance scale introduced from using the (incorrect) linear P-L relation is at most at a few per cent level. While this is small compared to other systematic errors, it will be important in future efforts to produce a Cepheid distance scale accurate to one per cent in order to remove degeneracies presented in CMB results. 
\end{abstract}


\section{Introduction}

The Cepheid period-luminosity (P-L) relation is a fundamental tool in cosmology because it plays a major role in the distance scale ladder. A well calibrated PL relation enables Cepheid distance measurements up to $\sim30$Mpc with the Hubble Space Telescope ($HST$). These Cepheid distances can then be used to calibrate a host of secondary distance indicators which can then, in turn, be used to measure Hubble's Constant that is free of the local Hubble flow \citep[see, e.g.,][]{fre01,sah01}. This has been done, for example, in the $H_0$ Key Project that obtained Cepheid distances to $\sim30$ $HST$ observed galaxies. It is also important in stellar pulsation/evolution studies. Empirical P-L relations also provide strong constraints to theoretical pulsation and evolution models. 

The current most widely used P-L relation is based on the Large Magellanic Cloud (LMC) Cepheids. This is primarily because there are many Cepheids discovered in the LMC. The LMC P-L relation has been assumed, and believed, to be linear in $\log(P)$ for a long time, where $P$ is the pulsation period in days. However this assumption is currently under challenge from several recent papers \citep{tam02,kan04,kan06,san04,nge05,nge06}. These studies present strong evidence that the LMC P-L relation may not be linear, and further, that the relation can be broken into two relations with a discontinuity at (or around) 10 days. In particular, a rigorous statistical test (the $F$-test) for analyzing the nonlinearity of the LMC P-L relation, using the $V$-band data from both of the OGLE (Optical Gravitational Lensing Experiment) and the MACHO (Massive Compact Halo Object) samples has returned a significant result: this is reported in \citet{kan04,kan06} and \citet{nge05}, respectively. The fact that the statistical test returns a significant result from two totally {\it independent} samples strongly indicates that the nonlinear LMC P-L relation is real and not due to artifacts such as photometric reductions and observational strategies.

Despite this, the evidence of the nonlinear LMC P-L relation is still a controversial issue and draws much skepticism within the astronomical community. However this does not rule out the possible existence of an intrinsic, but hard-to-detect nonlinear P-L relation. The purpose of this paper is to examine some of the mis-conceptions and clarify some of the issues regarding the detection of a nonlinear LMC P-L relation. In this paper the nonlinear P-L relation discussed is defined as two P-L relations with a discontinuity at/around a period of 10 days that separate the short and long period Cepheids. Hence short and long period Cepheids are those with period less and greater than 10 days, respectively. Justifications of the choice of a fiducial break period at 10 days are given in \citet{kan04}, \citet{san04}, \citet{nge05} and \citet{nge06}, and will not be repeated here. We note that \citet{nge05} also estimated this break period from the data. Other forms of the nonlinear P-L relation can be found in \citet{nge05}, however the study from \citet{nge06} suggested the two-lines relation is more appropriate to represent the nonlinearity. The confirmation of the break period at 10 days and the true form of the nonlinear P-L relation is difficult to verify with the observed data, due to the intrinsic dispersion of the P-L relation, and has to wait for more pulsational modeling. Some early attempts to explain theoretically and model the nonlinear P-L relation can be found in \citet{kan06} and \citet{mar05}. 

\section{Visualizing the Nonlinear P-L Relation}

One of the arguments against the idea of a nonlinear P-L relation is that the P-L relation plotted from observed data looks linear by human eye, therefore the relation should be linear. However, {\it the appearance of the linear P-L relation does not imply that the underlying P-L relation is indeed linear}. We demonstrate that it is difficult to distinguish these two P-L relations by eye with a simple experiment. We generate (i.e., simulation) two ``fake'' P-L relations to mimic the distribution of real Cepheids along the P-L regression: one is intrinsically linear and another one is intrinsically nonlinear.

For demonstration purposes, we use the $V$-band P-L relations given in \citet{san04}. The linear version of the P-L relation is: 

\begin{eqnarray}
V & = & -2.70 \log(P) - 1.49, 
\end{eqnarray}

\ni and the nonlinear version of the P-L relations are:

\begin{eqnarray}
V & = & -2.96 \log(P) - 1.34\ \mathrm{for} \ \log(P)<1.0, \\
V & = & -2.57 \log(P) - 1.63\ \mathrm{for} \ \log(P)>1.0.
\end{eqnarray}

\ni To obtain a realistic period distribution, we use the period distribution (in $\log[P]$) from the MACHO sample \citep{alc99} and the OGLE sample \citep{uda99b} as input to generate a period distribution that matches the observed distribution. It is well known that the Cepheid P-L relation has an intrinsic dispersion due to the finite width of the instability strip. We model this intrinsic dispersion as a Gaussian distribution with $\sigma=0.23$ mag centered at zero. In addition to the intrinsic dispersion, we also add a Gaussian error of $\sigma=0.05$ mag to represent photometric errors. These two Gaussian errors are then added in quadrature to the generated P-L relations using the above equations and our period distribution. 

\begin{figure}
\plotone{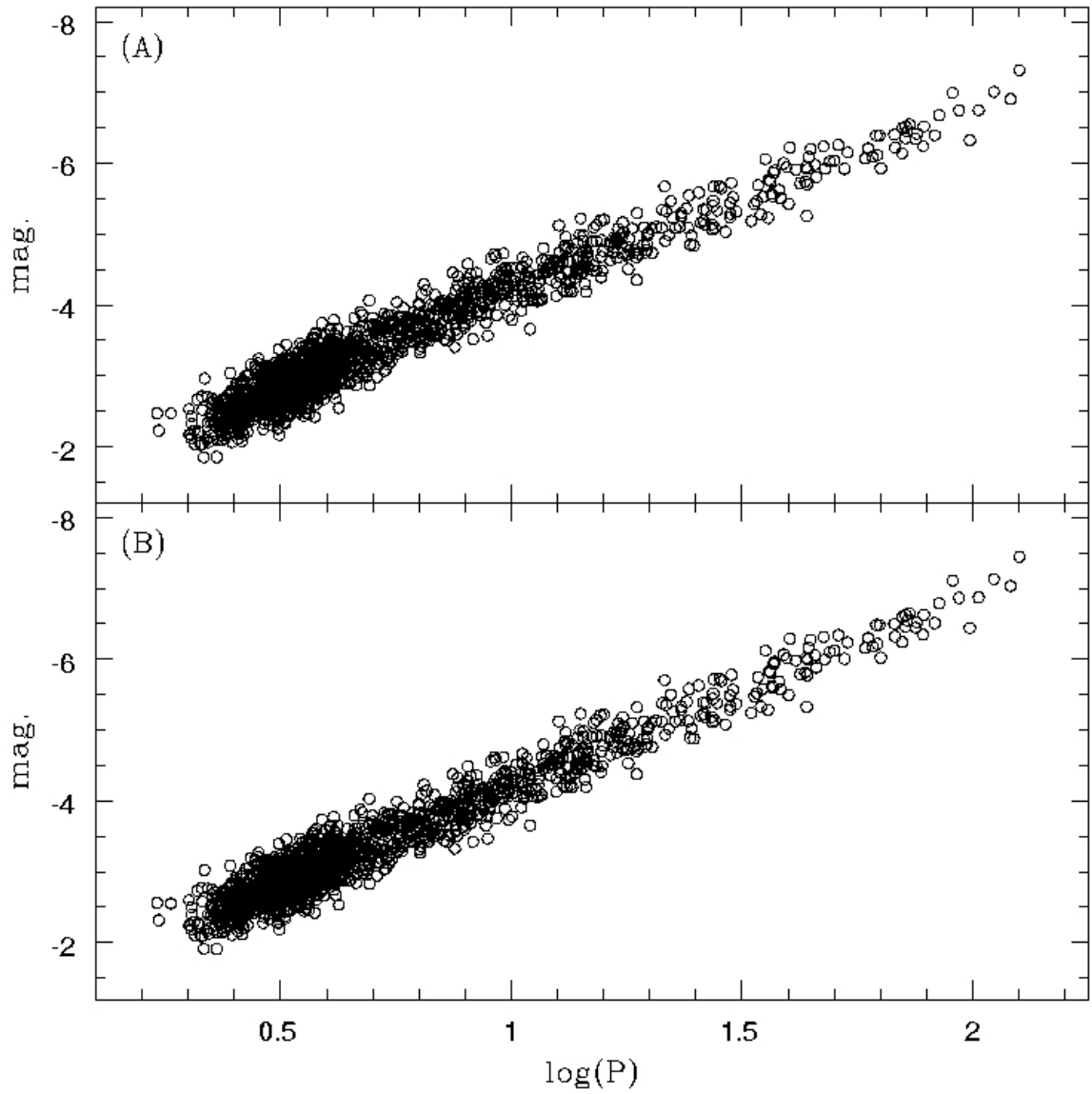}
\caption{The ``simulated'' P-L relations from 1500 Cepheids. Which of the P-L relation is intrinsic linear and which one of them is nonlinear? \label{figsimu1}}
\end{figure}

\begin{figure}
\plotone{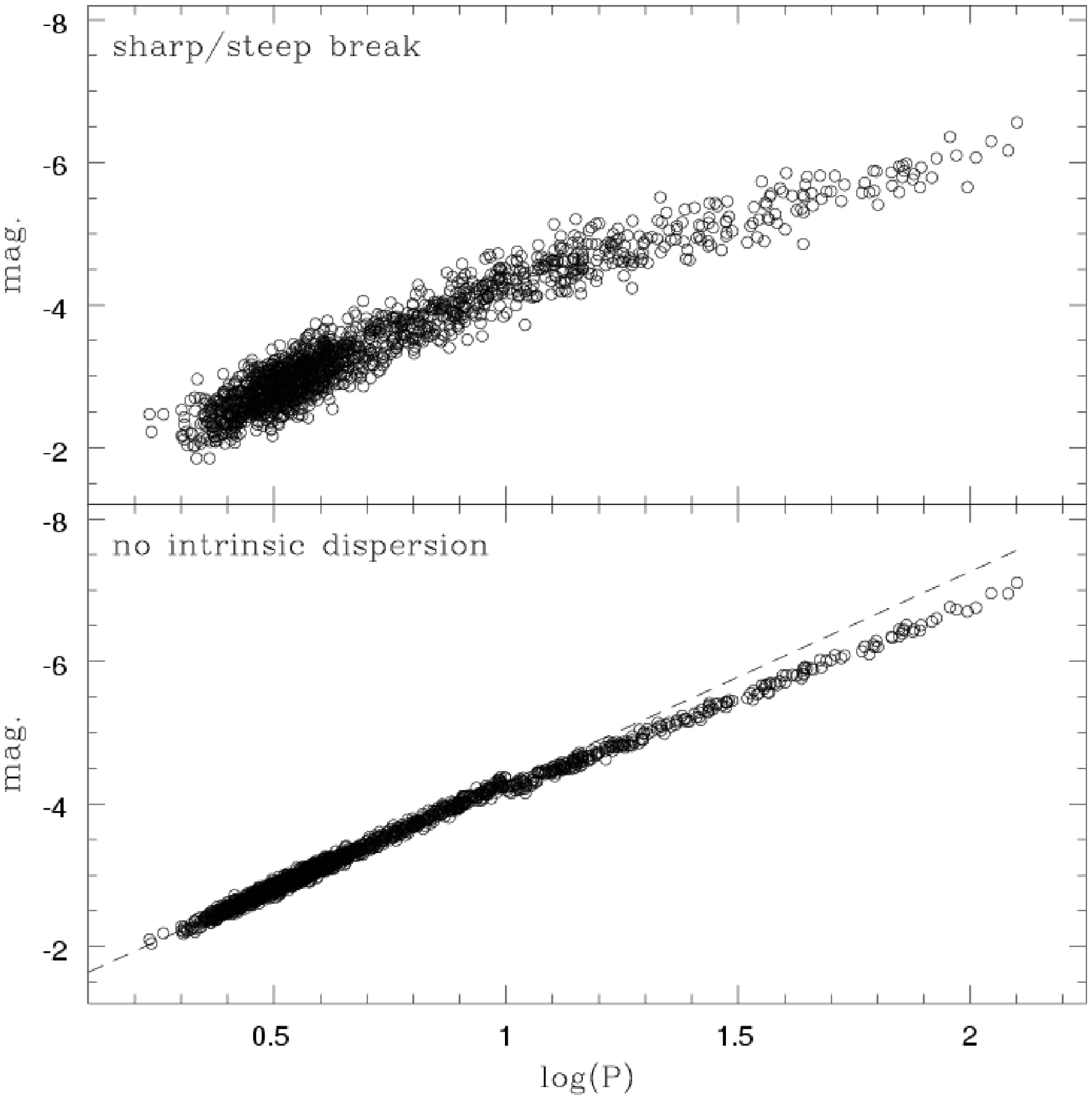}
\caption{{\it Top panel:} Simulated P-L relation with a sharp/steep break at 10 days. The long period relation used to generate this P-L relation is $V=-1.80\log(P)-2.50$ and the short period relation is same as in Figure \ref{figsimu1}. {\it Bottom panel:} simulated P-L relation without the intrinsic dispersion. A dashed line is drawn for the short period Cepheids and extended to the long period end to guide the eye. \label{figsimu2}}
\end{figure}

We plot out the simulated P-L relations in Figure \ref{figsimu1}. These are similar to the published LMC P-L relation \citep[e.g.][]{uda99a}. By looking at these P-L relations, it is very difficult to distinguish the nonlinear and linear P-L relations (see Appendix A for the answer). There are two reasons that make the nonlinear P-L relation difficult to visualize. Firstly, the existence of intrinsic dispersion dominates the total dispersion of the relation. Secondly, the nonlinearity of the P-L relation is shallow, or the break is not a sharp or steep break that will be easy to visualize. A shallow break of the P-L relation would be, for example, that the slope variation between the long and short period slopes is roughly $20$\% or less. To illustrate these effects we plot another two simulated P-L relations in Figure \ref{figsimu2}: one of them has a sharp/steep break and another one does not have intrinsic dispersion. The nonlinearity of the P-L relations in this figure is more apparent than the P-L relations given in Figure \ref{figsimu1}. 

\begin{figure}
\plotone{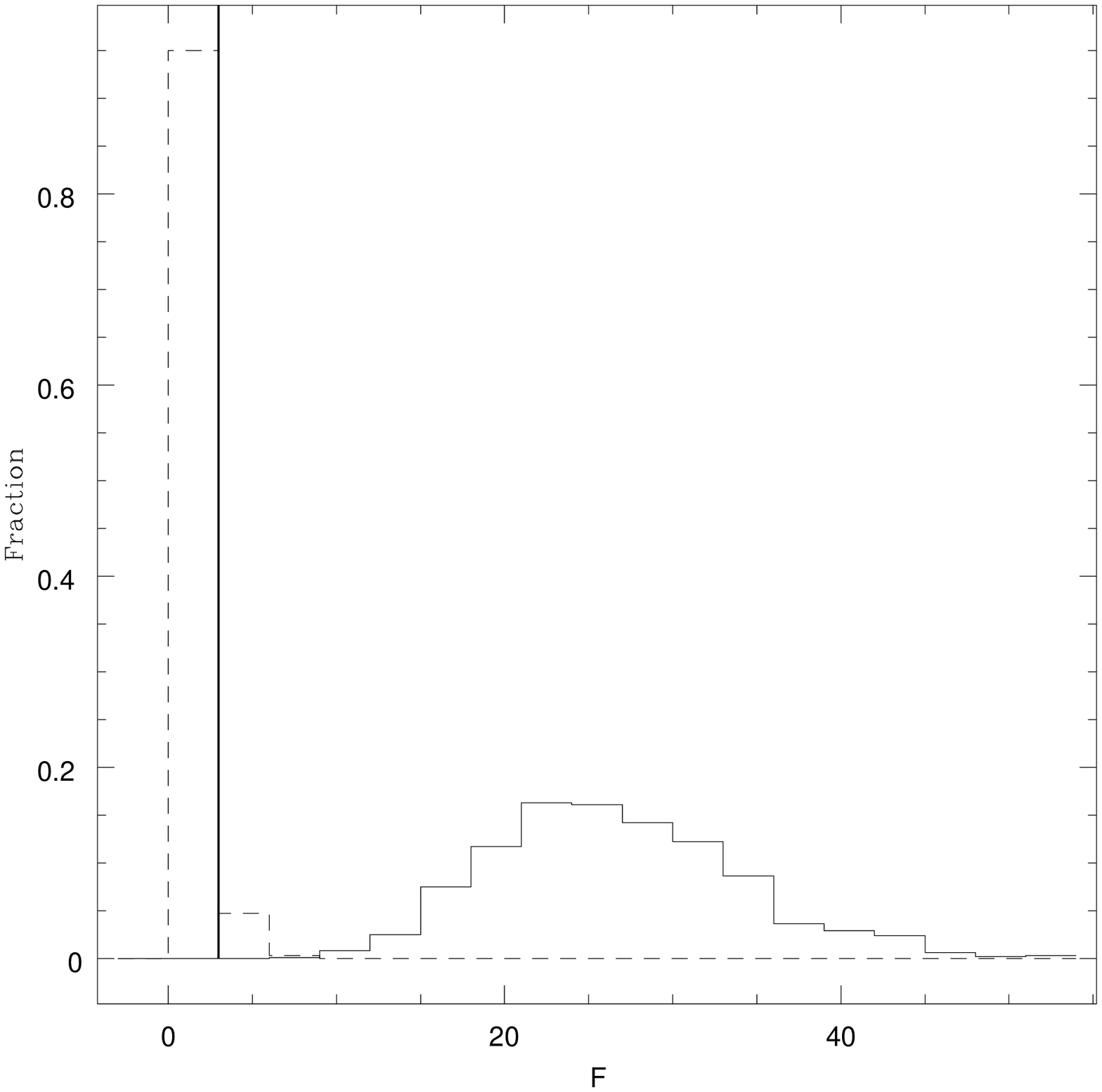}
\caption{The distributions of the values of $F$ from the $F$-test for simulated P-L relations that are intrinsically linear (dashed lines) and nonlinear (solid lines). For our simulated sample, $F>3$ indicates that the P-L relation is nonlinear with more than 95 per cent confidence level. Therefore a thick vertical line is drawn at $F=3$ and we choose the bin-size to be $3$ when building up these histograms. \label{figfhist}}
\end{figure}

Based on the above demonstration, how then can we be confident about detecting the existence of a nonlinearity in the P-L relation, if any? The answer is that we need to use some {\it statistical tests and/or analysis} to detect the hidden nonlinearity. One of the statistical tests that we employ in our analysis is the $F$-test which can be found in many statistical texts \citep[e.g.,][]{wei80}. The $F$-test has been applied to the OGLE data \citep{kan04,kan06} and the MACHO data \citep{nge05} and both datasets return a significant result for the nonlinearity for the $V$-band LMC P-L relation (see the $F$-test results in the Appendix A for the P-L relation in Figure \ref{figsimu1}). To demonstrate that the $F$-test can return a reliable result, we repeat the ``simulation'' that was done in Figure \ref{figsimu1} many times and build up the distribution of the $F$-test results. The resulting histograms for both of the intrinsically linear and nonlinear simulated P-L relations are shown in Figure \ref{figfhist}. It can be clearly seen from this figure that the $F$-test will return a significant (100 per cent of the time) and a non-significant (95 per cent of the time) result when the P-L relation is intrinsic nonlinear and linear, respectively. Other tests that maybe apply to test and detect the nonlinear P-L relation include, for example, the non-parametric regression as outlined in \citet{nge05}.

\section{Number of Long Period Cepheids and Sample Selections}

Currently perhaps the best LMC Cepheid samples are from the OGLE and MACHO projects. They are the "best" in terms of the number of Cepheids and the quality of the photometric data (for example, small photometric errors and large number of data points per light curves). Even though both samples show a nonlinear P-L relation as detected in \citet{tam02}, \citet{kan04} and \citet{nge05}, both samples lack Cepheids with $\log(P)>1.5$ (due to the saturation of CCD). The lack of longer period Cepheids, together with the (overall) smaller number of long period Cepheids, is another criticism against the detection of the nonlinear P-L relation. It has been suggested that the nonlinearity of the LMC P-L relation is due to the relatively small number of the long period Cepheids in the OGLE and MACHO samples, and this nonlinear feature will go away if more long period Cepheids are added to the samples. \citet{tam02} and \citet{kan04} using the OGLE data {\it alone} already suggested the LMC P-L relation is nonlinear. \citet{san04} tried to ameliorate the shortage of long period Cepheid by including additional Cepheids from literature to expand the OGLE sample. They found the nonlinear LMC P-L relation was still evident. The number of long period Cepheids will not be the same as the short period Cepheids. This is not a surprise, as argued in \citet{nge05}, because in general the mass is higher for longer period Cepheids. The consequence is that there are a fewer number of long period Cepheids (due to initial mass function) and the long period Cepheids will cross the instability in a shorter amount of time \citep{bon00}. These two effects have limited the number of long period Cepheids present in the LMC (or in any galaxy). 

Despite the fact that the $F$-test first used in \citet{kan04} {\it is} sensitive to both the number and nature of Cepheids on either side of the period cut, we will demonstrate that the sample selection in general will not affect the detection of a nonlinear LMC P-L relation. The published $V$-band mean magnitudes from various sources, as given below, will be used for this test because there is a vast amount of data available in the $V$-band. 

\begin{enumerate}
\item We downloaded the published OGLE data, the extinction corrected mean $V$ magnitudes and periods, from their website\footnote{http://bulge.princeton.edu/$\sim$ogle/ogle2/cepheids/query.html}. There are 771 Cepheids, but we removed those without $V$-band measurements. This left 762 in our ``OGLE'' sample. 

\item We add additional data from table 4 of \citet{seb02} to the OGLE data. We removed the entries in the Sebo dataset that are labeled with ``ogle'' to avoid duplicity with the OGLE Cepheids. We took the $E(B-V)$ values from \citet{per04} and/or \citet{san04}, if available, or set to $0.10$ (LMC mean extinction) otherwise for this sample to correct for the extinction. This sample is added to the OGLE sample given in item (1) above and called ``OGLE+SEBO''. 

\item We include the extinction corrected $V$-band data from \citet[][with 45 Cepheids]{lan94}, \citet[with 88 Cepheids]{cal91}, \citet[with 97 Cepheids]{san04} and \citet[with 53 Cepheids]{gie98}, separately, to our OGLE data in item (1), and they are called ``OGLE+LS94'', ``OGLE+CL91'', ``OGLE+STR04'' and ``OGLE+GFG98'', respectively.

\end{enumerate}

Following \citet{uda99a} and \citet{nge05}, Cepheids with $\log(P)<0.4$ were removed from the samples to avoid contamination from first overtone Cepheids. The plots of these samples (not shown) reveal that some of the obvious outliers should be removed. We use the sigma-clipping algorithm to remove these outliers \citep{uda99a}. First we fit a standard regression to all Cepheids in the samples and obtain the $\sigma$ ( = Root Mean Square, or total dispersion), then those outliers with a dispersion more than $X\sigma$ from the fitted regression line are removed. We adopt a novice value of $X=2.5$ in our test, which is also used by the OGLE team \citep{uda99a}. We then re-fit the regression to the short and long period Cepheids and apply the $F$-test as done in \citet{kan04} and \citet{nge05}. For our sample, $F\sim3$ at 95 per cent confidence level, hence if $F>3$ then the P-L relation is nonlinear. The results are summarized in Table \ref{tab1}. From this table it can be seen that the nonlinear P-L is clearly evident from the $F$-test using the OGLE data alone. Appending additional data from various sources to the OGLE data the nonlinearity of the LMC P-L relation is still preserved from the $F$-test. Including the Cepheids with period shorter than $\log(P)<0.4$ will not alter the detection of nonlinear P-L relation given in this table. A similar test is also done to the MACHO sample \citep{nge05}, as both the MACHO sample alone and the ``MACHO+SEBO'' samples combined have returned a significant $F$-test result. Therefore we believe that {\it the detection of the nonlinear LMC P-L relation is not due to the sample selection and the lack of longer period Cepheids in the samples}, as long as we have a good sample (with large number of Cepheids, see next section as well). 

\begin{deluxetable}{lccccccc}
\tabletypesize{\scriptsize}
\tablecaption{Testing the nonlinearity of the P-L relation with different samples.\label{tab1}}
\tablewidth{0pt}
\tablehead{
\colhead{Dataset} & \colhead{slope$_S$} & \colhead{ZP$_S$} & \colhead{$N_S$} & \colhead{slope$_L$} & \colhead{ZP$_L$} & \colhead{$N_L$} & \colhead{$F$}
}
\startdata
OGLE         & $-2.888\pm0.057$ & $17.152\pm0.036$ & 644 & $-2.467\pm0.258$ & $16.806\pm0.306$ & 54 & 6.70 \\
OGLE+Sebo    & $-2.836\pm0.054$ & $17.117\pm0.034$ & 708 & $-2.594\pm0.149$ & $16.961\pm0.175$ & 97 & 6.02 \\
OGLE+LS94    & $-2.883\pm0.055$ & $17.149\pm0.034$ & 656 & $-2.799\pm0.121$ & $17.205\pm0.158$ & 87 & 7.28 \\
OGLE+CL91    & $-2.850\pm0.053$ & $17.134\pm0.033$ & 681 & $-2.819\pm0.113$ & $17.241\pm0.146$ & 105& 7.24 \\
OGLE+STR04   & $-2.884\pm0.063$ & $17.159\pm0.040$ & 693 & $-2.560\pm0.110$ & $16.909\pm0.144$ & 116& 6.41 \\
OGLE+GFG98   & $-2.865\pm0.054$ & $17.140\pm0.034$ & 666 & $-2.762\pm0.134$ & $17.161\pm0.171$ & 85 & 6.36 
\enddata
\tablecomments{Subscripts $_S$ and $_L$ referred to short and long period Cepheids, respectively.}
\end{deluxetable}

\begin{figure}
\plotone{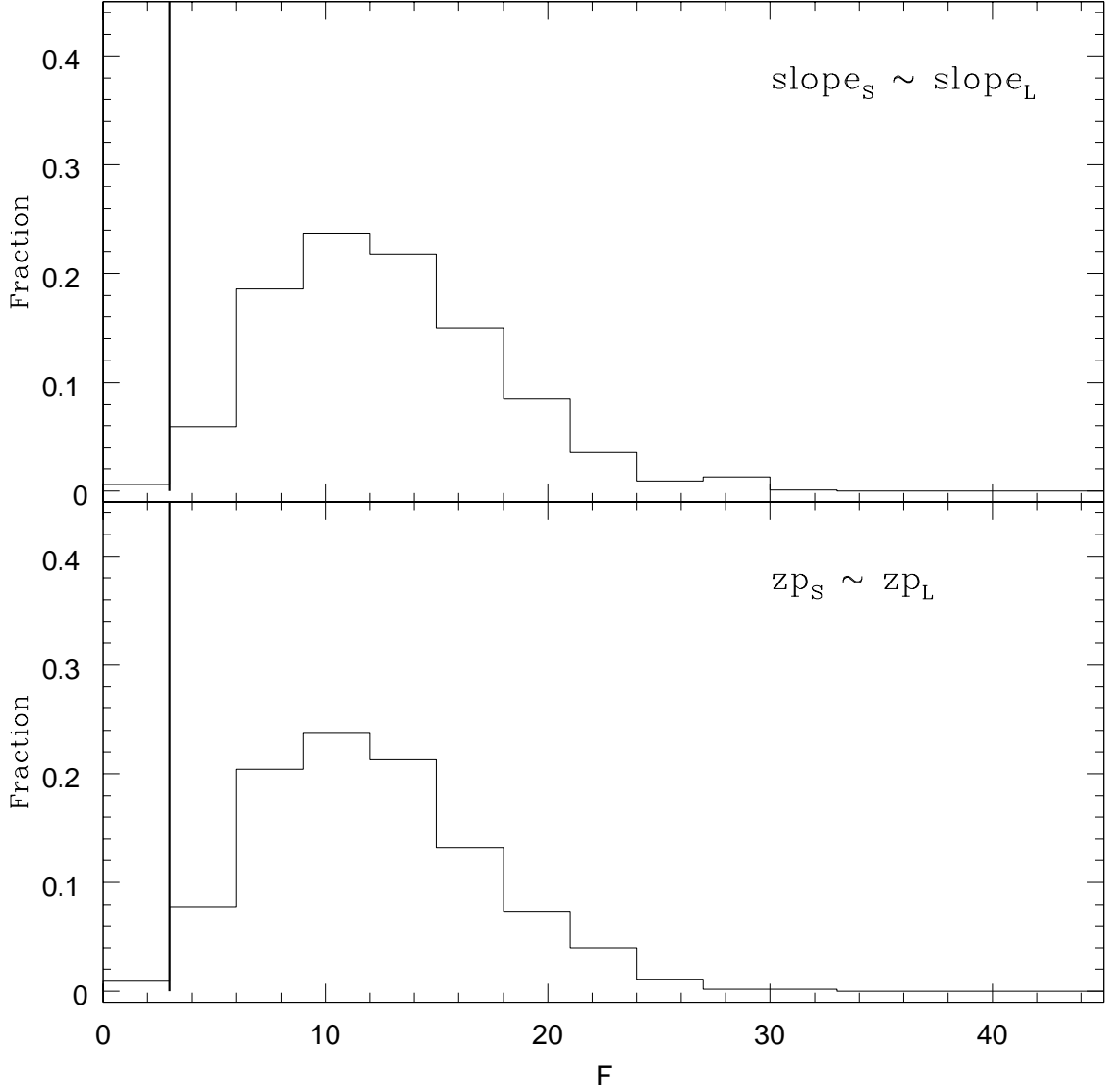}
\caption{Histograms for the $F$-test results when the slopes (top panel) and zero-points (bottom panel) are similar for the long and short period P-L relations in the ``simulations''. The vertical thick lines separate the linear ($F<3$) and nonlinear regimes. \label{fighist2}}
\end{figure}

It is worthwhile to point out that the P-L relation is believed to be linear if the slopes of the long and short period P-L relations are similar. Some cases of the similar slopes can be seen from Table \ref{tab1} (e.g., OGLE+CL91). However nonlinearity of the P-L relation will depend on both of the slopes and the zero-points, and the full form of nonlinear P-L relations is taken into account in the $F$-test. In Figure \ref{fighist2}, we show the histograms of the $F$-test results from the simulated P-L relations using the ``OGLE+CL91'' and ``OGLE+GFG98'' P-L relations given in Table \ref{tab1} as the input P-L relations. The former P-L relations have similar slopes while the latter P-L relations have similar zero-points. From the figure the nonlinear P-L relation is still detected in both cases. 

\section{Removing Outliers}

\begin{figure}
\plotone{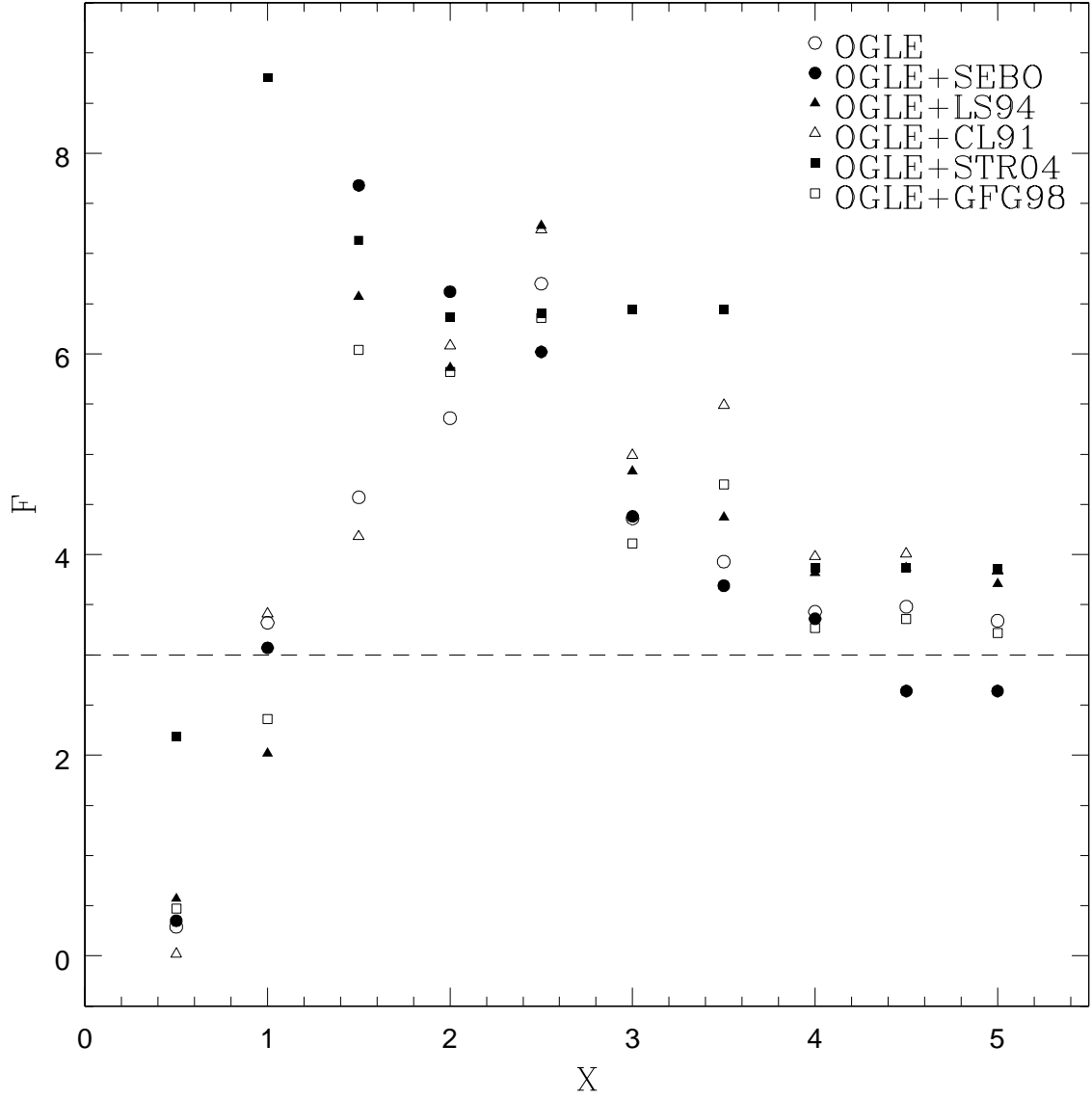}
\caption{Results of the $F$-test for the six samples with different adopted values of $X$ in the sigma-clipping algorithm. The dashed line is the case for $F=3$. \label{figoutlier}}
\end{figure}

In the previous section the outliers were removed using the sigma-clipping algorithm with $X=2.5$. In this section we will investigate the influence of outliers on the detection of the nonlinear P-L relation. Using the six samples as given in Table \ref{tab1} (again, Cepheids with $\log[P]<0.4$ are removed), we apply the sigma-clipping algorithm with 
\begin{eqnarray}
X=\{5.0,\ 4.5,\ 4.0,\ 3.5,\ 3.0,\ 2.5,\ 2.0,\ 1.5,\ 1.0,\ 0.5\} \nonumber
\end{eqnarray}
\ni to remove outliers and then calculate the corresponding $F$-values. The results are summarized in Figure \ref{figoutlier}. From this figure it can be seen that a general trend exists between the adopted value of $X$ and the resulting $F$-values for the LMC Cepheids. At high $X$, the samples show a marginal detection ($F\sim3$) of the nonlinearity of the P-L relation. As the adopted value of $X$ decreases to $\sim3$ and $\sim2$, the detection of the nonlinear P-L relation becomes significant.  When the adopted value of $X$ decreases to $\sim1$ or less, the P-L relation becomes linear as indicated by a non-significant $F$-test result ($F<3$). The cases for using large values of $X$ that did not show a nonlinear P-L relation can be understood as due to the influence of outliers, because these outliers will decrease the accuracy of the determined slopes and hence reduce the significance of the $F$-test \citep{nge05}. Adopting large value of $X$ in the sigma-clipping algorithm probably is not a good practice because some genuine outliers will not be removed (in other words, not many outliers will be removed). The cases for using small values of $X$ that do not show the nonlinearity of the P-L relation can be understood as follows: the intrinsic dispersion of the P-L relation is of the order of $\sim2\sigma$ and sigma-clipping with $X<1.5$ will remove many Cepheids, including those good Cepheids near the edges of the instability strip. This will make the regression tighter and tighter along the regression line. If nonlinearity {\it does} exist in the P-L relation within the intrinsic dispersion, the sigma-clipping with small $X$ will ``remove'' the signature of the nonlinear P-L relation and hence the $F$-test returns a non-significant result. Therefore, too high or too low of $X$ is not a good choice for rejecting the outliers. A novice choice of $X$ is $\sim2.5$ which we used in this paper as well as by the OGLE team.

Rejection of outliers within the samples, as we demonstrate here, is more critical to detect the nonlinear P-L relation in the LMC Cepheids, if any. Hence we are left with the following choices: (a) use a large value of $X$ (say $X>4$) in sigma-clipping algorithm that will include some obvious outliers in the sample and hence a marginal or no detection for the nonlinearity of the P-L relation; (b) use a small value of $X$ (say $X<1$) that removes a lot of the good Cepheids, as well as the outliers, and the nonlinear P-L relation will not be detected; or (c) use a reasonable value of $X$ (say $X\sim2.5$) to remove the outliers but the remaining sample will suggest the P-L relation is nonlinear. Choice (a) is not a good practice in astronomy as the obvious outliers will be included in the sample. Choice (b) is also not a good practice for obvious reason (why remove some good data together with the outliers?). Hence we left with choice (c) and conclude that the LMC P-L relation is nonlinear after a proper rejection of the outliers. Therefore we only have to {\it select a good sample that is free of (obvious) outliers, and the detection of the linear/nonlinear P-L relation will follow if the intrinsic P-L relation is linear/nonlinear}.

\section{Issues of Extinction}

Among the reasons that may cause the observed nonlinear P-L relation, extinction is believed to be the most likely candidate (assuming that the P-L relation is intrinsic linear). Therefore it is important to investigate and discuss the issues related to extinction further. It has been suggested that the $E(B-V)$ values given by the OGLE team are generally higher \citep[see, e.g.,][]{sub05}, therefore in this section we examine the LMC P-L relation using the extinction map provided by \citet{sub05}\footnote{Note that this extinction map only covers the bar region of the LMC, which is same as in the OGLE extinction map.}, who used the red clump stars along the line of sight to estimate the extinction values. Besides the lower extinction values, there is another difference between the extinction maps presented in \citet{sub05} and by the OGLE team: the spatial resolution of the \citet{sub05} extinction map, ($3.56\times3.56\ \mathrm{arcmin}^2$), is smaller than the OGLE extinction map ($14.23\times14.23\ \mathrm{arcmin}^2$). Given the locations of the OGLE Cepheids, the extinction map returns the corresponding $E(V-I)$ values and we can convert these values to $E(B-V)$ using $E(B-V)=E(V-I)/1.4$. However, there is a small fraction of the Cepheids located within the region without a reliable estimation of the $E(V-I)$ values \citep[see figure 4 in][]{sub05}. We therefore adopt/retain the original $E(B-V)$ values from the OGLE extinction map for these Cepheids. The $V$-band mean magnitudes are then corrected with the new extinction values. The $F$-test result for this OGLE sample (with $\log[P]<0.4$ removal and $X=2.5$ sigma-clipping algorithm) is $F=3.76$, which still indicates a nonlinear P-L relation. \citet{nge05} used a different extinction map from \citet{zar04} to the MACHO data and the subsequent $V$-band P-L relation is still nonlinear. Therefore we believe that the nonlinear detection of LMC P-L relation is independent of the extinction maps used.

\begin{figure}
\plotone{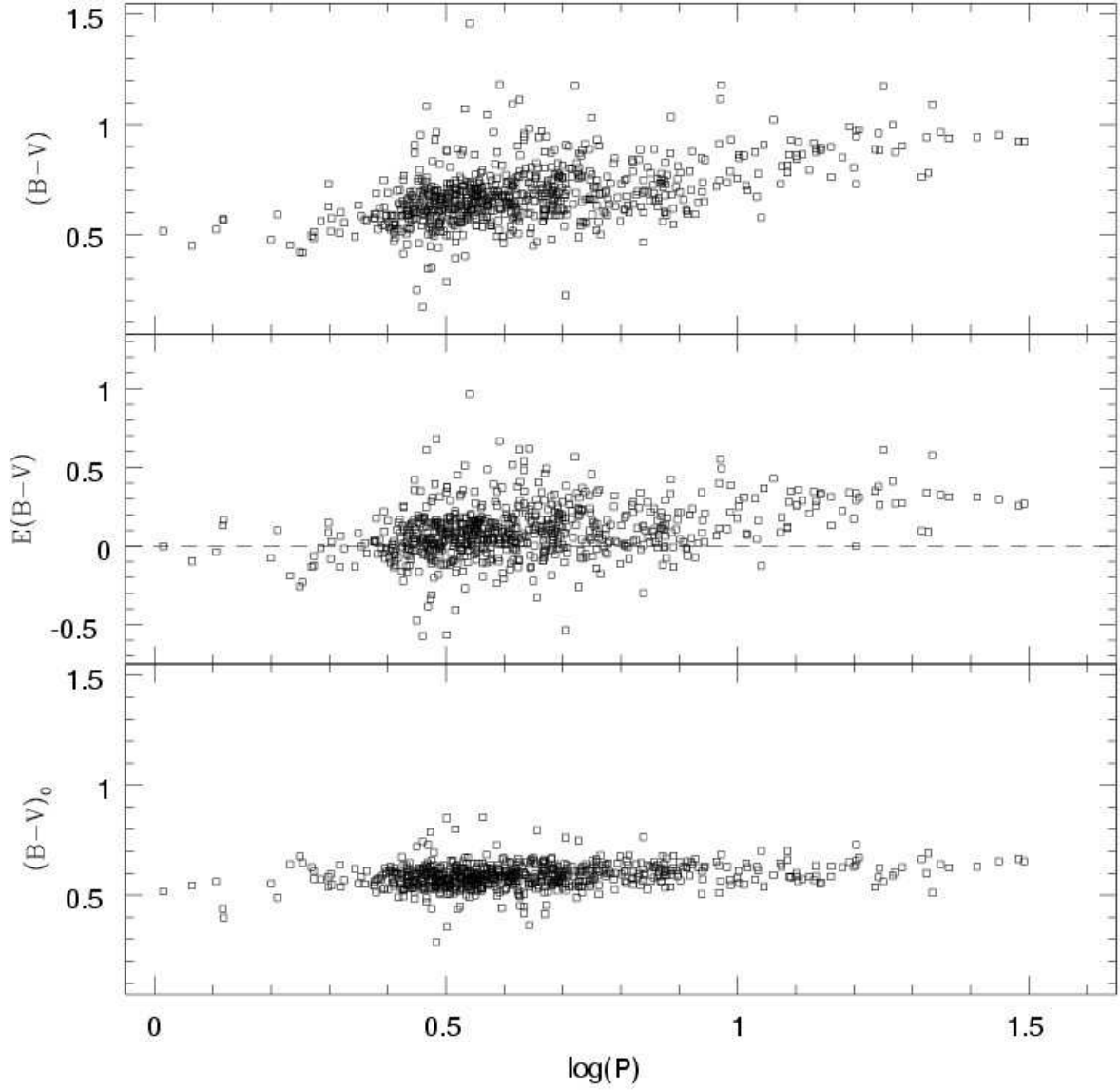}
\caption{{\it Top panel}: The unreddened $(B-V)$ P-C relation from the OGLE sample. {\it Middle panel}: The $E(B-V)$ values from \citet{mad82}'s relation as function of period. {\it Bottom panel}: The extinction corrected $(B-V)$ P-C relation, which is simply obtained from $(B-V)-E(B-V)$, i.e. top panel minus middle panel. \label{figbv}}
\end{figure}

To explain the observed nonlinear behavior of the LMC P-L (and P-C) relation as being due to extinction errors, the regression for the $E(B-V)$ values and $\log(P)$ for individual Cepheids will have the form of $E(B-V)=\alpha\log(P)+\beta$, with $\alpha$ being a positive value to counter the apparent nonlinear P-L relation. In addition, the extinction error hypothesis would not only have to explain the existence for such a relation but also explain why this relation exists only for the long period Cepheids. We examine such a period dependency of $E(B-V)$ values in the following two ways:

\begin{enumerate}

\item A regression can be fit to the OGLE Cepheids with $\log(P)$ and $E(B-V)$ values from the extinction maps and hence obtain a value for $\alpha$. Using the $E(B-V)$ values from the original OGLE extinction map, we obtained $\alpha=-0.011\pm0.004$ and $-0.014\pm0.005$ for all Cepheids and only the long period Cepheids, respectively. Using the \citet{sub05} extinction map, we have $\alpha=-0.015\pm0.006$ (for all Cepheids) and $-0.026\pm0.008$ (for long period Cepheids). The negative values of $\alpha$ are in contradiction with what is expected (a positive value) from the extinction hypothesis as described. In fact, some of the longer period Cepheids ($\log[P]>1.6$) have comparable or lower $E(B-V)$ values than the short period Cepheids \citep[see, e.g., table 1 of][]{san04}.

\item \citet{mad82} has provided a relation for $E(B-V)$ that include a period dependency in the relation: $E(B-V)=-0.26V-1.05\log(P)+1.84(B-V)+3.62$. Using the OGLE Cepheids with available $V$ magnitudes and $(B-V)$ colors, we can obtain the $E(B-V)$ values with this relation and hence correct for extinction. The $F$-test result for this extinction corrected $V$-band data is $F=0.79$. This implies that the P-L relation is linear if we use \citet{mad82}'s relation for the extinction correction. However there are three caveats for using \citet{mad82}'s method. First of all, the resulting slope for the $V$-band P-L relation is too steep, with $-3.615\pm0.064$ for all Cepheids in the sample ($-3.542\pm0.092$ and $-3.457\pm0.349$ for short and long period Cepheids, respectively). This is in direct contradiction to the observed $V$-band P-L slope that should be around $\sim -2.7$ to $\sim -2.8$. This range of slope is also consistent with theoretical predictions \citep[see, e.g.,][]{cap00}. Secondly, the unreddened $(B-V)$ P-C relation already displays a nonlinear feature, and this nonlinear feature will propagate to the $E(B-V)$-$\log(P)$ plot with \citet{mad82}'s relation (see the middle panel of Figure \ref{figbv}). As a result the extinction corrected $(B-V)$ P-C relation and hence the $V$-band P-L relation will become linear, as displayed in Figure \ref{figbv} for the $(B-V)$ P-C relation. In other words, the \citet{mad82}'s relation cannot be used to correct for extinction if the observed P-C relation is nonlinear. Finally, about one quarter of the subsequent $E(B-V)$ values are negative (see middle panel of Figure \ref{figbv}), which is physically unrealistic. Therefore, \citet{mad82}'s relation cannot be used to account for the nonlinear P-L and P-C relations as an extinction effect, even though the relation contains a period dependency. If this resolution is adopted, then the resulting LMC PL relation must be the one to be used for calibrating extra-galactic distances. Because the slope is so different in this case, this will mean a large change in the existing extra-galactic distance scale. 

\end{enumerate}

\begin{figure}
\plotone{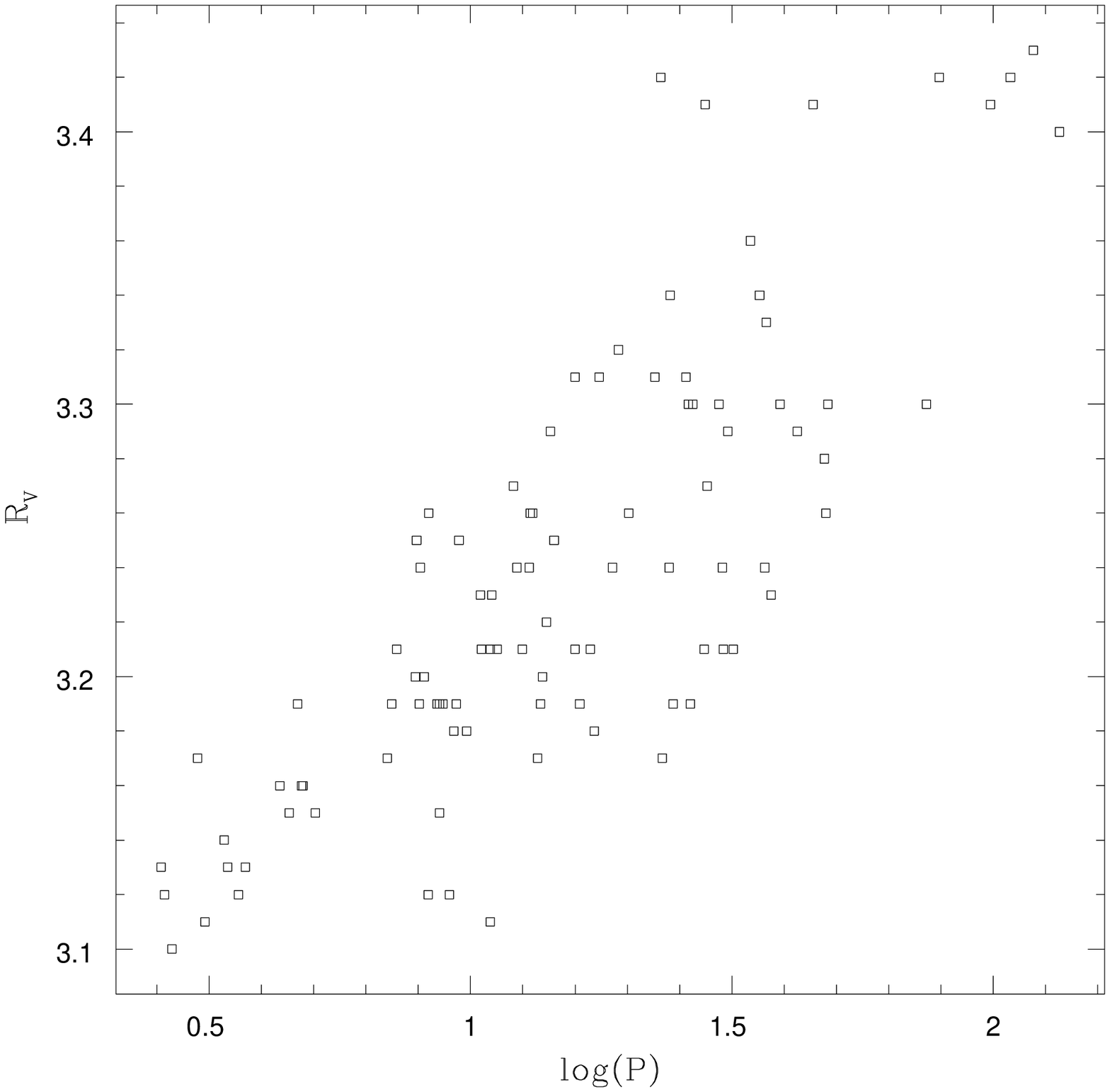}
\caption{Plot of the $R_V$ as a function of period using the data given in table 1 of \citet{san04}. The fitted regression is $R_V=0.165(\pm0.013)\log(P)+3.043(\pm0.016)$. \label{figrv}}
\end{figure}

\ni We hence believe that the extinction hypothesis with a period dependency on the $E(B-V)$ values may not be able to account the for observed nonlinear LMC P-L and P-C relations. Since the extinction correction is $A_V=R_VE(B-V)$, could the total-to-selection extinction coefficient $R_V$ be responsible to the nonlinear LMC P-L relation? In this paper, we adopt a constant value of $R_V=3.24$ from \citet{uda99b}. Using the $\sim90$ $R_V$ values given in table 1 of \citet{san04}, we plot $R_V$ as a function of period in Figure \ref{figrv}. The figure has suggested that there is a correlation between $R_V$ and $\log(P)$. However, if we adopt such a period dependency of $R_V$ and apply this to our OGLE data (with the OGLE extinction map), we still obtain a significant detection of nonlinearity of the P-L relation with $F=6.74$. Therefore a period dependency of $R_V$ still cannot explain the observed nonlinear LMC P-L relation. 

Another piece of evidence against extinction errors causing the observed nonlinear P-L relation is the P-C relation at maximum light. The observed P-C relation for the Galactic and LMC long period Cepheids is shown to be flat at the maximum light \citep{cod47,kan04}. There is a sound physical reason behind the flatness of the P-C relation at the maximum light \citep{sim93,kan04a,kan06}: the interaction of hydrogen ionization front (HIF) and photosphere at the maximum light. As period increases, Cepheids can only get cooler or stay roughly in the same temperature range (i.e. the P-C relation is flat) at maximum light. If an additional amount of extinction is needed to make the P-C relation linear at the mean light, the the same amount of extinction would force the Cepheids to become hotter as period increases at maximum light. This is in serious contradiction with the pulsation theories and observations. In fact, some researchers have used the flatness of the P-C relation at the maximum light to derive the extinction values \citep{sim93,fer94}. Furthermore, the multi-phase study of the LMC P-L and P-C relations implies that the LMC P-L and P-C relations are nonlinear at most of the phases over the pulsation cycle, especially at phases near $0.8$ \citep{nge06}. Based on the above arguments and the arguments presented in \citet{kan04}, \citet{san04}, \citet{nge05} and \citet{kan06}, we believe that {\it extinction errors are not responsible for the observed nonlinear P-L (and P-C) relation in the LMC Cepheids} from the current available extinction studies.

\section{The Period-Color Relation is also Nonlinear}

The P-L relation and the period-color (P-C) relation for Cepheid variables are not independent of each other. \citet{mad91} have given a thorough review for the physics behind the Cepheid P-L and P-C relations. At mean light in the optical bands, a nonlinear P-L relation will imply that the P-C relation is also nonlinear and vice versa. The maximum light P-C relation shows a slightly different behavior, see \citet{nge06}. Compared to the P-L relation, the nonlinearity of the P-C relation is compelling and easier to visualize \citep[see the P-C plots in][]{tam02,kan04,san04,nge05,kan06}. However the evidence for the nonlinear LMC P-C relation is largely ignored when the detection of nonlinear P-L relation is criticized. The large scatter at given period in the P-C relation is given as an example of possible errors in extinction. However, we state again that if the extinction errors are great, they should also affect the PC relation at maximum light but the P-C relation at maximum light is flat as predicted by theory. The $F$-test applied to the P-C relation has already been discussed in \citet{kan04,kan06} and \citet{nge05} and will not be repeated here. The results again strongly support the nonlinear P-C relation. Unless new evidence and/or new theory emerges to show that the Cepheid P-L and P-C relations should be totally independent, {\it the linear or nonlinear properties for both of the optical P-L and P-C relations cannot be neglected, and a nonlinear P-C relation would imply a nonlinear P-L relation}.

\section{The Nonlinearity of the LMC P-L Relations in Different Pass-Bands}

There is a mis-conception that if the P-L relation is nonlinear in the optical ($BVRI$) bands, then the same degree of nonlinearity of the P-L relation should also be present in the near infrared ($JHK$) bands (or at least the nonlinear $J$-band P-L relation would be easily visible), and vice versa. \citet{per04} presented the $JHK$-band P-L relations from $\sim92$ LMC Cepheids and found that these P-L relations are linear. Since the $JHK$-band P-L relations are linear, then, based on the above mis-conception, an argument against the finding of the nonlinear P-L relation is that the optical P-L relation should also be linear. As suggested and argued in \citet{nge05}, the main reason that \citet{per04} found the $JHK$-band P-L relations to be linear is because there are not enough short period Cepheids ($\sim15$) in their sample, as compared to over 600 short period Cepheids found in the OGLE and MACHO samples. As a consequence the $F$-test will not return a significant result for detecting the nonlinear P-L relation. Furthermore, \citet{nge05} has used the MACHO and 2MASS (Two Micron All Sky Survey) data to study the nonlinearity of the P-L relation in $VRJHK$ bands. Their results imply that the LMC P-L relation in nonlinear in the $VRJH$-band but linear in the $K$-band. 

\begin{figure}
\plotone{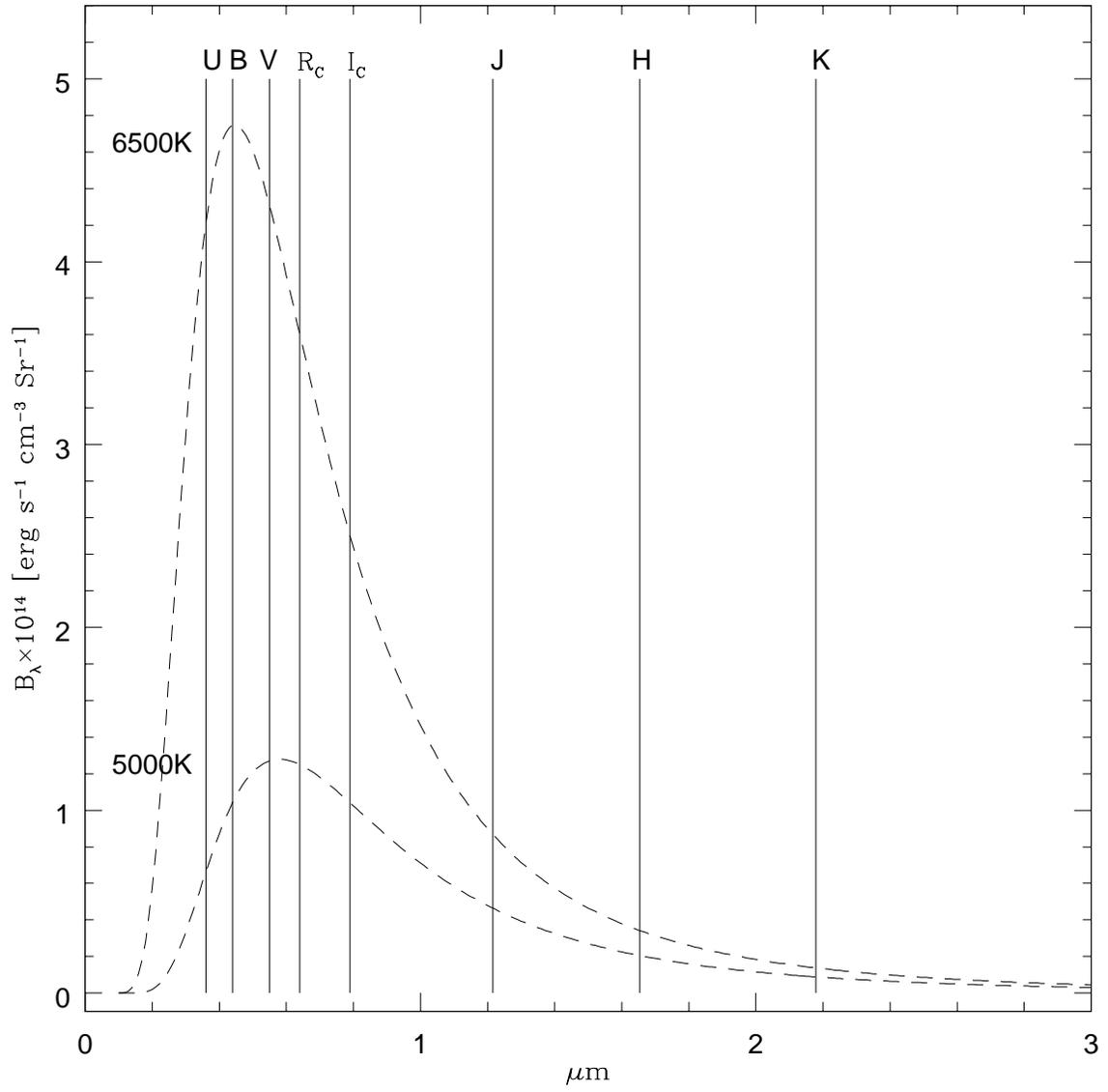}
\caption{The black-body curves at two temperatures that bracket the typical temperature variation of the Cepheid variables. Vertical lines indicate roughly the locations of different pass-bands. \label{figbb}}
\end{figure}

It is well-known that, at least in the optical, the temperature variation dominates the luminosity variation of Cepheid variables \citep[see, e.g.,][]{cox80}, and the range of the temperature variation is around $\sim 4500K$ to $\sim6500K$. In Figure \ref{figbb}, we plot the black-body curves for two temperatures that roughly bracket typical temperature variations in Cepheids. We use vertical lines to indicate the locations of different pass-bands. From this figure it is clear that the luminosity variation is greatest in the optical bands, and this variation decreases toward the infrared bands. Since the LMC (optical) P-C relation is also found to be nonlinear and there is no obvious reason that the Cepheid period-radius (P-R) relation should not be linear \citep[see, e.g.][]{bon98,gie98}, we assume that the nonlinear P-L relation is due to the temperature behavior (i.e. the nonlinear P-C relation) of the LMC Cepheids. Then from Figure \ref{figbb}, it is straight forward to see why the optical band P-L relations will exhibit a higher degree of nonlinearity than the $J$- and $H$-band P-L relation, and also why the $K$-band P-L relation is expected to be linear. Therefore, {\it the degree of nonlinearity is not the same in all bands, as the optical band P-L relation will show a higher degree of nonlinearity than the near infrared bands ($J$- and $H$-band)}. On the other hand, the results found in \citet{nge05}, i.e. the nonlinear $VRJH$-band P-L relations and a linear $K$-band P-L relation, suggested that the temperature behavior of the LMC Cepheids could play an important role in determining the nonlinearity of the LMC P-L relation. \citet{kan06} has proposed a possible mechanism, the interaction of the HIF with the photosphere, that can alter the temperature behavior of the LMC Cepheid during a pulsating cycle and lead to the observed nonlinear P-L relation.

\section{Nonlinear LMC P-L Relation versus Universality of the P-L Relations}

It is well-known that the zero-point of the P-L relation could be metallicity dependent, i.e. the metal-rich Cepheids are intrinsically brighter than metal-poor Cepheids at a given period \citep[see, for examples,][]{ken98,sto04}. In contrast, the slope of the P-L relation is assumed and considered to be metallicity independent, i.e. universal, and hence the P-L relation can be calibrated with large numbers of well observed LMC Cepheids. This assumption of a universal P-L relation has greatly simplified the use of P-L relations in distance scale applications. Since the Galactic P-L relation is linear, then, based on the universal assumption, the LMC P-L relation should also be linear with the same slope as found in the Galactic P-L relation. This is another argument against the detection of the nonlinear P-L relation for the LMC Cepheids. However recent observational and empirical studies imply that there are three different point of views regarding the metallicity dependency of the slope for the Cepheid P-L relation: 

\begin{enumerate}

\item  No metallicity dependency: The slope of the Cepheid P-L relation is totally independent of the metallicity and its value is universal. Papers supporting this point of view include, for example, \citet{gie98}, \citet{uda01} and \citet{gie05}\footnote{Note that the P-L slopes for Galactic Cepheids found in \citet{gie05} with their new $p$-factor relation (to convert the radial velocity to pulsational velocity) is shallower than the ``old'' slopes. This is in disagreement with the steeper P-L slopes from the Galactic Cepheids with open cluster distances alone, as found in \citet{tam03}, \citet{nge04} and \citet{san04}. Using the 12 Galactic Cepheids given in table 5 of \citet{gie05}, the slopes of the $V$-band P-L relation with distances from the open clusters, the old $p$-factor relation and the new $p$-factor relation are $-2.97\pm0.25$, $-2.97\pm0.22$ and $-2.79\pm0.22$, respectively. It is true that these three slopes are consistent with each other to within the (large) errors (due to the small number of Cepheids in the sample). However these slopes give a rough idea that the new $p$-factor relation will produce a shallower slope and hence the slope is closer to the LMC P-L slope. In contrast, the slope from the old $p$-factor relation is (almost) identical to the slope obtained with the open cluster distances. See \citet{kan06} and footnote 8 of \citet{sah06} for further discussion.}. 

\item  Evidence of metallicity dependency: Studies from \citet{tam03}, \citet{nge04} and \citet{san04} provided evidence that the slope of the Galactic P-L relation is statistically significantly different to that in the LMC, hence the slope of the P-L relation can not be universal, at least at the LMC and Galactic metallicity. 

\item  Evidence of metallicity dependency up to a certain value: \citet{rom05} plotted the $V$-band residuals from the \citet{uda99a} P-L relation as function of iron abundance for a number of Galactic and Magellanic Cloud Cepheids. Their result implied that Cepheids get fainter as metallicity increases (linearly) until the Solar metallicity, where a turn-over or a plateau occurs.

\end{enumerate}

\ni However, because the number of Galactic Cepheids (less than 100) that are applied to define the Galactic P-L relation is much smaller than the LMC counterparts, it is still premature to confirm which of the above points of view represents reality. On the other hand, recent theoretical studies that incorporate the latest input physics (e.g., treatments of time-dependent convection) have suggested the Cepheid P-L relation is not universal \citep[see, e.g.,][]{ali99,bon99,cap00,fio02,mar05}, in contrast to the older results \citep[e.g.,][]{ibe84,sto88}. Given the recent observational/empirical and the theoretical results, the assumption of universal P-L relation is not fully justified. Therefore the {\it detection of nonlinear LMC P-L relation may not be ruled out based on the argument of a universal P-L relation alone}. The issue of universality of the P-L relation will be fully addressed in the future when better data for more Cepheids are available. 

In fact, the nonlinearity of the LMC P-L relation provides another evidence against the assumption of universal P-L relation since the observed Galactic P-L relation is shown to be linear \citep{nge04}. The theoretical LMC P-L relations broken at 10 days from \citet{mar05} are consistent with the observed nonlinear LMC P-L relations, while the theoretical Galactic P-L relation is shown to be less ``quadratic'' \citep[i.e., linear, see][]{bon99,mar05}. \citet{kan06} also provided preliminary reasons why the P-C (and hence the P-L) relations for Galactic and SMC (Small Magellanic Cloud) Cepheids are linear and for the LMC Cepheids are nonlinear in terms of the HIF-photosphere interaction. This interaction, to a large extent, depends on the metallicity. Further, \citet{nge06} studied the multi-phase P-C relations for the Galactic, LMC and SMC Cepheids and found that the P-C relations are different in these three galaxies at some phases. This leads to different P-C relations for Galactic, LMC and SMC Cepheids seen at mean light as found in \citet{tam03,kan04} and \citet{san04}. Therefore the P-L relation may not be universal because the P-L and P-C relations are not totally independent of each other. 

\section{Conclusion}

In this paper we have examine and discuss various aspects and issues concerning the nonlinear LMC P-L relation arising from recent studies. In summary, (a) we find that it is difficult to visualize the nonlinear P-L relation and statistical tests are needed to detect it; (b) we provide evidence that sample selection, long period Cepheids, outlier removals and extinction errors are unlikely to be responsible for the detection of the nonlinear P-L relation; and (c) we argue the existence of nonlinear P-L relation from the perspective of nonlinear P-C relation and the non-universality of the P-L relation. Combining the results from this paper and the results from previous studies, especially the nonlinear study on the MACHO data \citep{nge05}, there are strong indications that the {\it observed} LMC P-L relation is indeed nonlinear. Then the logical conclusions based on this are either one of the following or a combination of both:

\begin{enumerate}
\item  The nonlinear feature of the LMC P-L relation is intrinsic and real, and there are some physical reasons that are associated with the pulsational properties of Cepheids and cause the observed nonlinear P-L relation. We referred these as the ``internal reasons'' and further study in stellar structure, pulsation and evolution is crucial to model and explain the nonlinear P-L relation. For example, \citet{kan06} explain the nonlinear P-L relation as being due to the HIF-photosphere interaction, and metallicity plays an important role in determine the different behavior for such interaction. Theoretical P-L relations constructed from the LMC models are also in good agreement with the observed nonlinear P-L relation \citep{mar05}.

\item There are some hidden/additional factors that may cause the apparent nonlinear feature in the LMC data, and we refer these as the ``external reasons''. These hidden/additional factors, if they exist and are found to be responsible for the nonlinear P-L relation, need to be taken into account when establishing a P-L relation. Perhaps the treatment of extinction is incomplete and/or inaccurate in the literature and in our study. Using the interferometric technique, \citet{ker06} found that the circumstellar envelope (CSE) could exist in a long period Galactic Cepheids, l Car. The existence of CSE could potentially complicate the extinction correction for Cepheids. However, if we assume that CSE exists in both of the Galactic and LMC Cepheids and it causes the observed nonlinear LMC P-L relation, then the {\it observed} Galactic P-L relation should also be nonlinear. This is in contradiction with the current observed Galactic P-L relation which is linear (because the Galactic P-C relation is also linear, see, e.g., \citealt{kan04}). 
\end{enumerate}

\ni Nevertheless, detailed investigations of these two aspects are beyond the scope of this paper. They are the two main future research directions for explaining and understanding the observed nonlinear LMC P-L relation. Certainly, the nonlinearity of the LMC P-L relation poses a new problem in the study of Cepheid variables.

\subsection{Distance Scale Application}

Since the LMC P-L relation, which has been assumed to be linear, has been extensively applied in the distance scale studies in the past, the detection of nonlinear LMC P-L relation could imply a ``disaster'' for the past distance scale works. In particular, does the nonlinear P-L relation invalidate previous distance scale results? \citet{nge05w} has looked at this problem closely, and the authors found that if the Cepheid distance is obtained using the Wesenheit function, a linear combination of the P-L and P-C relation, then the typical difference in the distance modulus with a linear and a nonlinear P-L relation is $\sim \pm0.03$ mag, or at $\sim$1.5 per cent level. This is because the Wesenheit function is linear: the nonlinear P-L and P-C relations cancel each other out \citep{nge05w}. This finding is further supported by the study of \citet{nge06a}, who compared the Hubble constant obtained from the Type Ia supernovae that are calibrated with both the linear and nonlinear LMC P-L relations. The difference in the Hubble constant is still at the $\sim$1.5 per cent (or less) level. Therefore, {\it the nonlinear P-L relation will not seriously invalidate previous distance scale results at a few per cent level}. It is also not a surprise that \citet{pie06} found a similar and consistent distance to IC 1613 when using the linear and the nonlinear P-L relation. 

However, from the present study and the previous studies, there is growing evidence that the LMC P-L relation could be intrinsically nonlinear. Then the correct form of the (nonlinear) LMC P-L relation should be applied in the future distance scale studies \citep[see also][]{kan03,sah06,san06}. In this sense, the linear P-L relation is an approximation to the intrinsically nonlinear P-L relation. As in all other science, a simple assumption and approximation is applied before a more realistic but complicated relation is developed and/or discovered. \citet{nge05w} and \citet{nge06a} have also discussed the importance of using the (correct) nonlinear LMC P-L relation in distance scale studies. This is because using the (incorrect) linear P-L relation will introduce an additional error of $\sim \pm0.03$ mag to the systematic error of distance scale and/or the Hubble constant, which is important in the era of precision cosmology to reduce the total systematic error. It has been known that there is a degeneracy between the total density of the universe ($\Omega_{tot}$) and the Hubble constant \citep[see, e.g.,][]{teg04}, and a highly accurate measurement of the Hubble constant would help to break this degeneracy. The current uncertainty of the Hubble constant through the distance ladder measurement is at $\sim10$ per cent level \citep[see, e.g.,][]{fre01}, with systematic errors contributed from various sources (for example, the error in the LMC distance is still one of the largest systematic error for the Hubble constant). In the near future, when these systematic errors are refined and improved to $\sim1.5$ per cent level, which are comparable to the error introduced from the linear versus nonlinear P-L relation, the importance of eliminating the additional systematic error from the linear and nonlinear P-L relation cannot be neglected.

\acknowledgements

We thank C. D. Laney for providing the data presented in \citet{cal91}. We thank the anonymous for useful discussions and suggestions. This research was supported in part by NASA through the American Astronomical Society's Small Research Grant Program.

\appendix

\section{Answer to Figure \ref{figsimu1}}

The P-L relation in Figure \ref{figsimu1}(A) is intrinsic nonlinear and the P-L relation in Figure \ref{figsimu1}(B) is intrinsic linear. The $F$-values from the $F$-test for these two generated P-L relations are $35.0$ and $0.59$, respectively.

\end{document}